\documentclass[apj]{emulateapj}
\usepackage{apjfonts}
\usepackage{hyperref}
\usepackage{graphicx}

\newcommand{\mbh}{$M_{\rm BH}$}
\newcommand{\msigma}{$M_{\rm BH}-\sigma_{\star}$}

\shorttitle{AGN Black Hole Mass Database}
\shortauthors{Bentz \& Katz}

\received{}
\accepted{}

\begin{document}

\title{The AGN Black Hole Mass Database}

\author{ Misty~C.~Bentz\altaffilmark{1},
Sarah~Katz\altaffilmark{2,3}
}

\altaffiltext{1}{Department of Physics and Astronomy,
		 Georgia State University,
		 Atlanta, GA 30303, USA;
                 bentz@astro.gsu.edu}
\altaffiltext{2}{North Springs Charter High School,
                 7447 Roswell Rd,
                 Atlanta, GA 30328, USA}
\altaffiltext{3}{University of California,
                 Santa Barbara, CA 93106, USA}

\begin{abstract}

The AGN Black Hole Mass Database is a compilation of all published
spectroscopic reverberation-mapping studies of active galaxies.  We
have created a public web interface, where users may get the most
up-to-date black hole masses from reverberation mapping for any
particular active galactic nucleus (AGN), as well as obtain the
individual measurements upon which the masses are based and the
appropriate references.  While the database currently focuses on the
measurements necessary for black hole mass determinations, we also
plan to expand it in the future to include additional useful
information, such as host-galaxy characteristics.  New reverberation
mapping results will also be incorporated into the database as they
are published in peer-refereed journals.

\end{abstract}


\section{Introduction}

Variability has long been recognized as one of the defining
characteristics of active galactic nuclei (AGNs)
\citep{matthews63,smith63} and was also one of the key observational
properties that led to our current physical description of AGNs as
accreting supermassive black holes \citep{salpeter64}.  While the
nuclear activity is interpreted as a clear signature of a supermassive
black hole, AGNs are generally too distant for even the largest
telescopes to spatially resolve the gravitational influence of the
black hole and determine its mass with current technology.  Instead,
AGN black hole masses are most often directly determined through time
resolution of the AGN variability, rather than spatial resolution, in
a technique known as reverberation mapping \citep{blandford82}.

Reverberation mapping measures the time delay between changes in the
continuum emission (likely arising from the accretion disk) and the
response to these changes in the broad emission lines (arising from
the photoionized broad line region, BLR).  The time delay is
effectively the mean light travel time between the two regions and
thus a measure of distance at spatially-unresolvable scales ($\sim
0.01$\,pc in typical nearby Seyferts).  As the accretion disk is
generally expected to be much smaller than the BLR, the time delay
between continuum fluctuations and the response of a specific broad
emission line gives the responsivity-weighted mean radius of the BLR
for that particular emission species.  

One of the key discoveries from the first reverberation mapping
campaigns was that the time delay (or BLR size) was smaller by a
factor of $\sim 10$ than previously expected from
photoionization models \citep{peterson85}.  Early on, it was also
discovered that, as expected, different emission lines have different
mean time delays, evidencing ionization stratification within the BLR
gas \citep{clavel91,peterson91,dietrich93,maoz93}.  More highly
ionized species have shorter time delays relative to more neutral
species, and their line widths are larger, evidencing larger
line-of-sight gas velocities.  In fact, a virial relationship is
observed between time delays and line widths of different emission
species in a single object: time delays are inversely proportional to
the square of the line widths (e.g.,
\citealt{peterson99,peterson00,onken02}).  Combining the time delay
measurement with the line width therefore gives a measure of the black
hole mass, albeit modulo a scaling factor that encompasses the unknown
geometry and dynamics of the gas in the broad line region
\citep{wandel99b}.

As our knowledge of AGN physics has evolved, so have the observing
requirements for reverberation campaigns and the techniques used to
analyze them.  \citet{peterson04} undertook a massive homogeneous
reanalysis of all the existing reverberation datasets.  One result
from this reanalysis was that many of the early datasets were severely
undersampled in their time resolution.  While not surprising given the
original expectation that the broad line region was much larger, this
finding suggested that the time delays, and therefore black hole
masses, from these datasets were unreliable.  Since 2004, there have
been many new reverberation campaigns, some specifically designed to
replace early campaigns with poor time sampling, and others to extend
the sample to include new objects.  As a result, for those astronomers
not closely linked to the reverberation mapping community, it can be
confusing or unclear where to find the best determination of the black
hole mass for any specific AGN, or whether such a measurement even
exists.

In response, we undertook to compile all the existing reverberation
measurements in a single database with a homogeneous format and to
create a simple web interface for use by the community.  The result is
described in this manuscript.

\section{The Reverberation Sample}

The large investment of time and resources required for reverberation
mapping campaigns leads to the current moderate size of the sample.
There are 63 AGNs with successful reverberation campaigns in the
published literature, and the sample continues to evolve with time.
Most of these AGNs are located in the nearby Universe ($z<0.1$) and
are therefore some of the most well-studied AGNs at all wavelengths.
However, a handful of AGNs at higher redshifts also have reverberation
measurements, with the highest redshift currently held by S5\,0836+71
at $z=2.172$ \citep{kaspi07}.

There are two reasons why the sample is heavily skewed towards nearby
AGNs: (1) the large amount of observing time required for a
reverberation campaign generally limits the observations to small- and
medium-aperture telescopes in the 1.0-m to 4.0-m class, but the
necessity of high signal-to-noise spectra ($S/N \approx 100$) imposes a
relatively bright limit on the apparent magnitude of the target AGNs;
and (2) the observed time delay is a function of the intrinsic
luminosity of the AGN (e.g., \citealt{bentz13}) as well as its
distance from us: very luminous quasars have long time delays that are
exacerbated by time dilation as the photons traverse our expanding
Universe.  The observed time delay may be several years, leading to
the necessity for decade-long monitoring campaigns (e.g.,
\citealt{kaspi00,kaspi07}).  Bright nearby Seyferts may be
successfully monitored over a single observing season.  Resource
constraints, as well as student, postdoc, and junior faculty
timelines, therefore generally favor studies of nearby AGNs.

All of the AGNs in the Database are Type I or broad-lined AGNs,
because the broad emission lines are the main focus of any
spectroscopic reverberation mapping campaign.  The objects span the
range of observed Type I activity, including such well-studied AGNs as
NGC\,4395 (Seyfert 1.8), NGC\,4748 a.k.a.\ Mrk\,766 (narrow-line
Seyfert 1), NGC\,4151 (the prototypical Seyfert 1), 3C\,390.3
(double-peaked very broad-lined and radio loud AGN) and PG\,1226+023
a.k.a.\ 3C\,273 (flat-spectrum radio quasar).  The host galaxies also
show a broad diversity, ranging from the bulgeless disk galaxy of
NGC\,4395, to typical massive spirals, both barred and unbarred, to
elliptical galaxies.  Many of the AGN hosts show signs of recent or
ongoing merger activity (e.g., the tidal tail of Arp\,151 or the
ongoing merger of NGC\,3227 with NGC\,3226), but on the other hand,
many do not.

Finally, while the vast majority of reverberation measurements in the
literature focus on the optical regime, specifically the region around
H$\beta$ $\lambda 4861$, among the full sample there are measurements
for every strong broad emission line between Ly$\alpha$ $\lambda 1216$
and H$\alpha$ $\lambda 6563$, inclusive.  In particular, many objects
have reverberation measurements for rest-frame ultraviolet emission
lines from {\it International Ultraviolet Explorer} campaigns that
were carried out in the 1980s and 1990s.  The Database includes all
published measurements for all emission lines with reliable time lag
measurements.  We have not, however, included broad-band or
photometric reverberation mapping results (including X-ray bands) in
this compilation, as these measurements do not generally translate
into a direct mass constraint.

\section{The Database}

The AGN Black Hole Mass Database is a SQL (Structured Query Language)
database hosted at Georgia State University in downtown Atlanta, GA.
It contains all AGNs with published spectroscopic reverberation
mapping results published in refereed journals.  At the time of
writing this manuscript, there are 62 AGNs in the main database.  For
each of these AGNs, we have compiled the following general information
(where available):
\begin{itemize}
\item object name;
\item common alternate names;
\item coordinates of right ascension, declination, and redshift;
\item AGN activity classification;
\item {\it Hubble Space Telescope} optical (medium $V$) image of the
  host galaxy \footnote{For clarity and ease of comparing host-galaxy
    features between objects, we do not include non-{\it HST} images
    nor do we include images with different central wavelengths.  Only
    a handful of AGNs have no relevant image and no plans to acquire one in
    the near future.};
\item luminosity distance and angular diameter distance assuming a cosmology of 
      $H_0 = 71$\,km\,s$^{-1}$\,Mpc$^{-1}$, $\Omega_{M} = 0.30$, and 
      $\Omega_{\Lambda} = 0.70$.
\end{itemize}
Several ongoing studies are working to measure redshift-independent
distances to the AGNs in this sample, and we plan to incorporate those
results in the database when they are published.

In addition to the general information we have compiled for each AGN,
we have included the following information (where available) from the
reverberation mapping campaigns targeting each object:
\begin{itemize}
\item each broad emission line with published reverberation measurements;
\item Julian date range of the reverberation campaign; 
\item emission-line time delay measurement (in the rest-frame) relative to 
      the continuum as determined from: 
      (1) the peak of the cross-correlation function, $\tau_{\rm peak}$, 
      (2) the centroid of the cross-correlation function around the peak 
          above some threshold value, $\tau_{\rm cent}$, and 
      (3) the maximum likelihood from {\tt JAVELIN} \citep{zu11} based on a 
          damped random walk model of the continuum light curve and a 
          reprocessed emission-line light curve, $\tau_{\rm JAV}$;
\item emission line width in the root-mean-square residual spectrum
  (which highlights the variable part of the spectrum) as determined
  from: 
       (1) the full-width at half maximum, FWHM, and 
       (2) the second moment of the line profile, $\sigma_{\rm line}$;
\item references for the reverberation campaign and the time delay and 
      line width measurements;
\item the mean luminosity of the AGN at rest-frame
  5100\,\AA\ throughout the campaign, corrected for the contribution
  from host-galaxy starlight;
\item reference for the AGN luminosity.
\end{itemize}
The black hole masses are derived from a combination of the
measurements recorded for each object, and so they are calculated on
the fly as explained in the next section.

The successful measurement of a reverberation time delay requires not
only careful planning, but also a bit of luck.  AGNs that were
strongly variable in the past may not be strongly variable while they
are being monitored (cf.\ the monitoring results for Mrk\,290 as
described by \citealt{bentz09} and \citealt{denney09}).  Therefore, in
some cases, campaigns were published that were unable to measure a
reliable time delay between continuum and emission line variability
(e.g., Ark\,564, \citealt{shemmer01}).  Furthermore, some measurements
that were published in one study were reanalyzed (usually with updated
techniques) and found to be unconstrained.  Examples of this situation
include many of the famous Seyfert galaxies among the NGC sample,
e.g., NGC\,3227, NGC\,4051, NGC\,4151. The first monitoring datasets
for these objects, in particular, were prone to undersampling given
that they were among the earliest reverberation targets and the
observations were designed under the expectation that the broad line
region was much more extended, and therefore that the observed time
delays should be longer and require less fine sampling.  Datasets such
as we have described above have been listed in the database as
``excluded results'' with links to the appropriate manuscripts.  We
adopted the general approach to also list in ``excluded results'' any
datasets which have all reported emission-line time delay measurements
consistent with zero delay, as this is generally a sign of
undersampling in the monitoring cadence.  Datasets that contain
multiple emission-line measurements where only some (not all) of the
time delays are reported as being consistent with zero (e.g.,
\ion{He}{2} as measured for Mrk\,50 by \citealt{barth11}) are fully
included in the main database.  We believe this to be a fairly
conservative level of quality flagging, and we leave more careful
consideration of the quality of individual datasets to the user and
his/her best judgement.

In short, at this time, the database is a summary of 95 published
articles in peer-refereed journals with the results obtained for 74
individual AGNs.  Twelve of these AGNs only have ``excluded results''
at this time. For the 62 AGNs with included datasets, there are 226
sets of reverberation measurements tabulated in the database.
These are the basis for the determination of the black hole masses.

\section{Black Hole Masses}

Black hole masses are determined as:
\begin{equation}
M_{\rm BH} = f \frac{RV^2}{G}
\end{equation}
\noindent where $R$ is taken to be $c \tau_{\rm cent}$ and $V$ is
taken to be $\sigma_{\rm line}$, following \citet{peterson04}, and $c$
and $G$ are the speed of light and the gravitational constant,
respectively.  While $\tau_{\rm JAV}$ is often more tightly
constrained than $\tau_{\rm cent}$, $\tau_{\rm cent}$ is reported for
significantly more datasets (209 versus 119).  The factor $f$ is a
dimensionless scaling factor of order unity that encapsulates the
unknown geometry and dynamics of the broad line region gas. Typically,
the population-averaged value of $f$, denoted $\langle f \rangle$, is
determined as the multiplicative factor needed by the reverberation
masses to bring the \msigma\ relationship for AGNs into general
agreement with that of the dynamical sample of quiescent galaxies
(e.g., \citealt{mcconnell13}).  Values for $\langle f \rangle$ range
from $5.5 \pm 1.8$ \citep{onken04} to $2.8 \pm 0.6$ \citep{graham11},
depending on the objects included and the specific measurements
adopted for both samples (dynamical and reverberation).  We adopt the
\citet{grier13} value of $\langle f \rangle = 4.3$ as the default
value for determination of \mbh, but allow the user to choose other
values and recalculate the masses through the web interface (detailed
in the next section).  The uncertainties on the masses are based on
the combination of measurement uncertainties for $\tau_{\rm cent}$ and
$\sigma_{\rm line}$ only.

Many objects in the sample have measurements for multiple emission
lines, and/or multiple measurements of the same emission line.  We
adopt the weighted average of the masses derived from the measurements
of all emission lines as the default value of \mbh\ for each object,
although we also provide the mass based only on measurements of
H$\beta$ $\lambda 4861$.  Users who wish to determine the mass in some
other way, for example using $\tau_{\rm peak}$ and FWHM for \ion{C}{4}
only, will be able to do so with the information provided.  We caution
the user that, in general, other combinations of time delay and line
width measurements will require the adoption of other values of
$\langle f \rangle$ (e.g., \citealt{collin06}).

In addition to the derived reverberation masses, we also include other
direct mass measurements for the small number of objects where they
are available. In particular, NGC\,3227 and NGC\,4151 both have
\mbh\ determined from stellar dynamical modeling and from
spatially-resolved gas dynamical modeling \citep{davies06, hicks08,
  onken14}.  A handful of other objects --- Arp\,151, Mrk\,50,
Mrk\,1310, NGC\,5548, NGC\,6814, and SBS\,1116+583A --- have
\mbh\ derived directly from dynamical modeling of high quality
reverberation spectroscopic datasets \citep{pancoast12,pancoast14}.
As more such measurements become available in the literature, they
will be incorporated into the database.

\section{The Web Interface}

The information in the database is provided freely to the community
through the web interface, which is located at
\url{http://www.astro.gsu.edu/AGNmass}.

\begin{figure}
\plotone{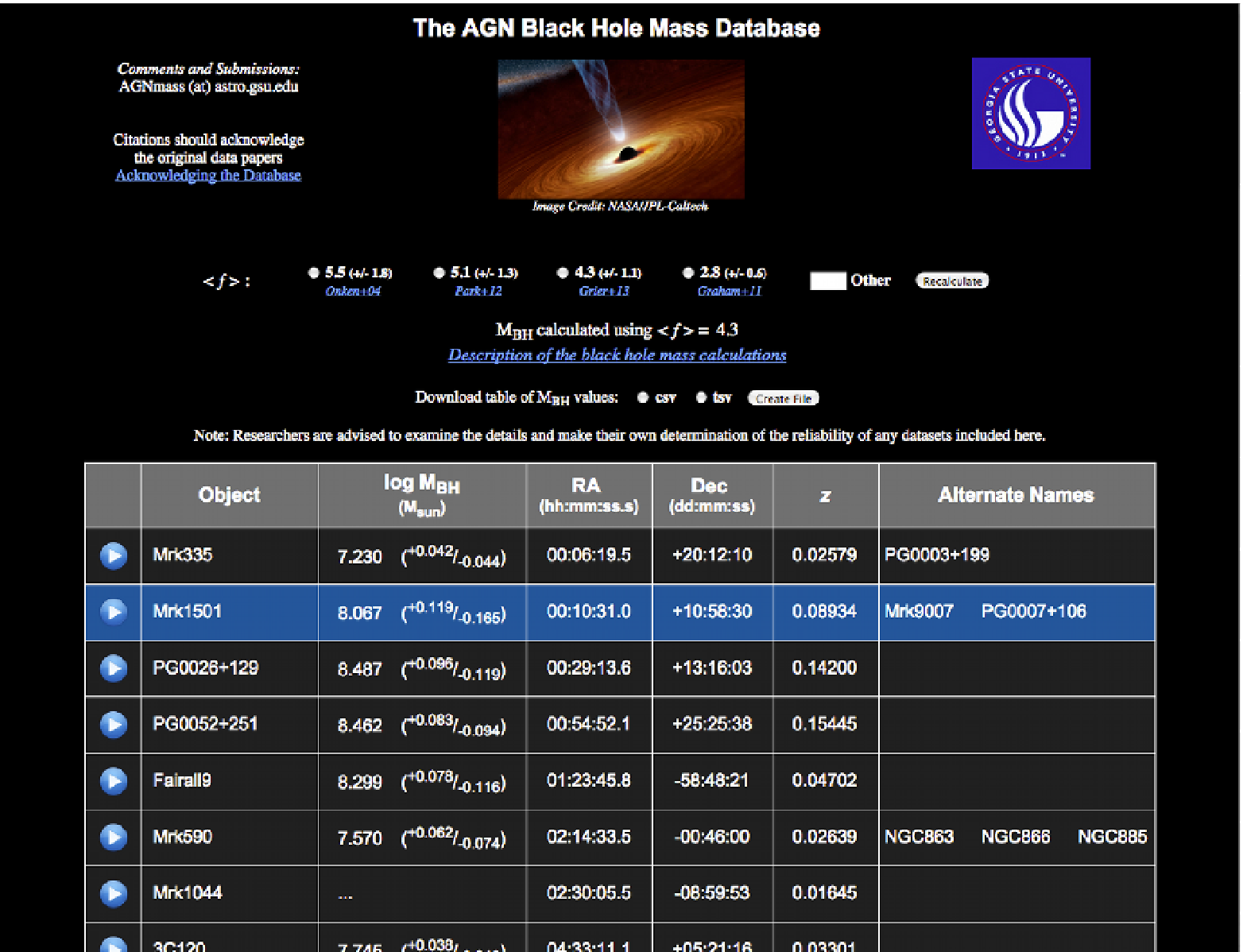}
\caption{The front page of the AGN Black Hole Mass
  Database web interface.  The main component of the front page is the
  object table with the black hole masses.  Also included are options
  for the user to change the adopted value of $\langle f \rangle$ in
  the mass determination, an option to create and download a file with
  either comma-separated or tab-separated values for the object table,
  and links to the detailed information for each object in the far
  left column of the object table.  There are also two anchors that
  will take the user to the bottom of the page where s/he can find
  information on the black hole mass calculations and how to
  acknowledge use of the database.}
\label{fig:front}
\end{figure}

Figure~\ref{fig:front} shows the front page that the user will
encounter when visiting the interface to the database.  All objects
included in the database are listed with their default black hole mass
calculations, coordinates, and common alternate names, so the user can
quickly identify whether an object of interest is included (or not).
Should the user prefer to adopt a different value of $\langle f
\rangle$, buttons are provided at the top allowing all masses to be
quickly recalculated with other commonly-adopted scaling factors,
including a field where the user may type any value they wish to use.
A short explanation of the mass calculations is provided at the bottom
of the page, and is a reiteration of the information described in this
manuscript in $\S 4$.  Furthermore, the user may download the object
table and black hole masses with the user's current adopted value of
$\langle f \rangle$ as either a comma-separated values (csv) or
tab-separated values (tsv) file.

If the user has javascript enabled in their browser, each row in the
object table will be highlighted when it is moused over to help with
object selection, as shown in Figure~\ref{fig:front} for Mrk\,1501.
Detailed information for each object, including a table of all
reverberation measurements, can be viewed by clicking on the button at
the left of each row in the object table.

\begin{figure}
\plotone{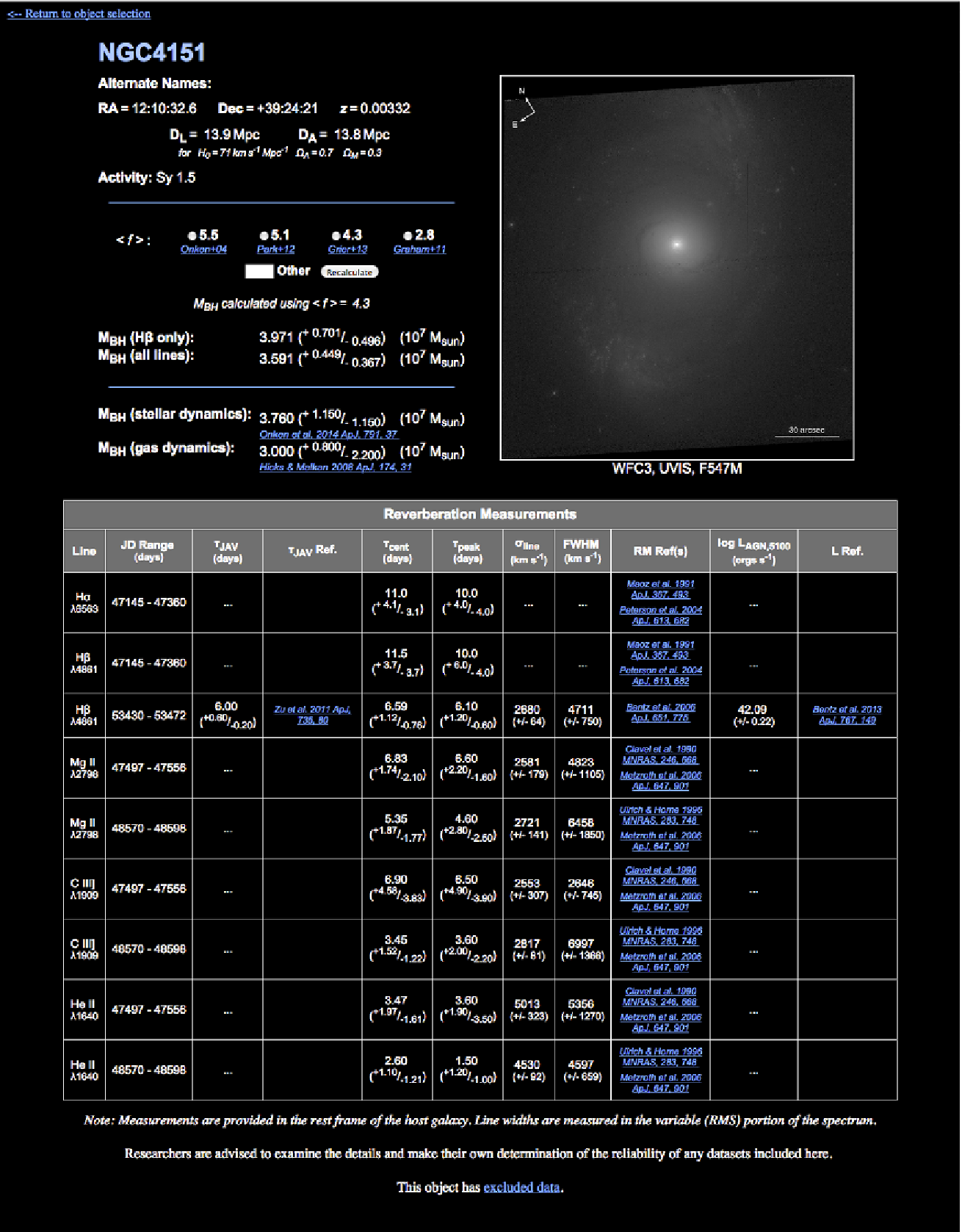}
\caption{The detailed page for an individual object, in
  this case, NGC\,4151.  The main component of the page is the
  reverberation mesurements table, with all measured values of time
  delay and line width for all emission lines. The black hole mass is
  again listed, with options for the user to change the adopted value
  of $\langle f \rangle$. Black hole masses determined only from
  H$\beta$ measurements or from another method (such as stellar or gas
  dynamical modeling) are also included for comparison.  If there are
  any excluded data for the object, a link will appear at the bottom
  of the page, such as is seen for NGC\,4151.}
\label{fig:details}
\end{figure}

Figure~\ref{fig:details} shows the page for an individual object with
all the details currently recorded in the AGN Black Hole Mass
Database.  This includes the information that was provided on the main
page in the object table, as well as the host-galaxy image, a table of
reverberation measurements, and the other information described in $\S
3$ and $\S 4$ of this manuscript.  The mass is again provided based on
all measurements of all emission lines, with a default value of
$\langle f \rangle = 4.3$, but is also provided based only on
measurements of H$\beta$.  The user is again able to recalculate the
mass with different adopted values of $\langle f \rangle$.

\begin{figure}
\plotone{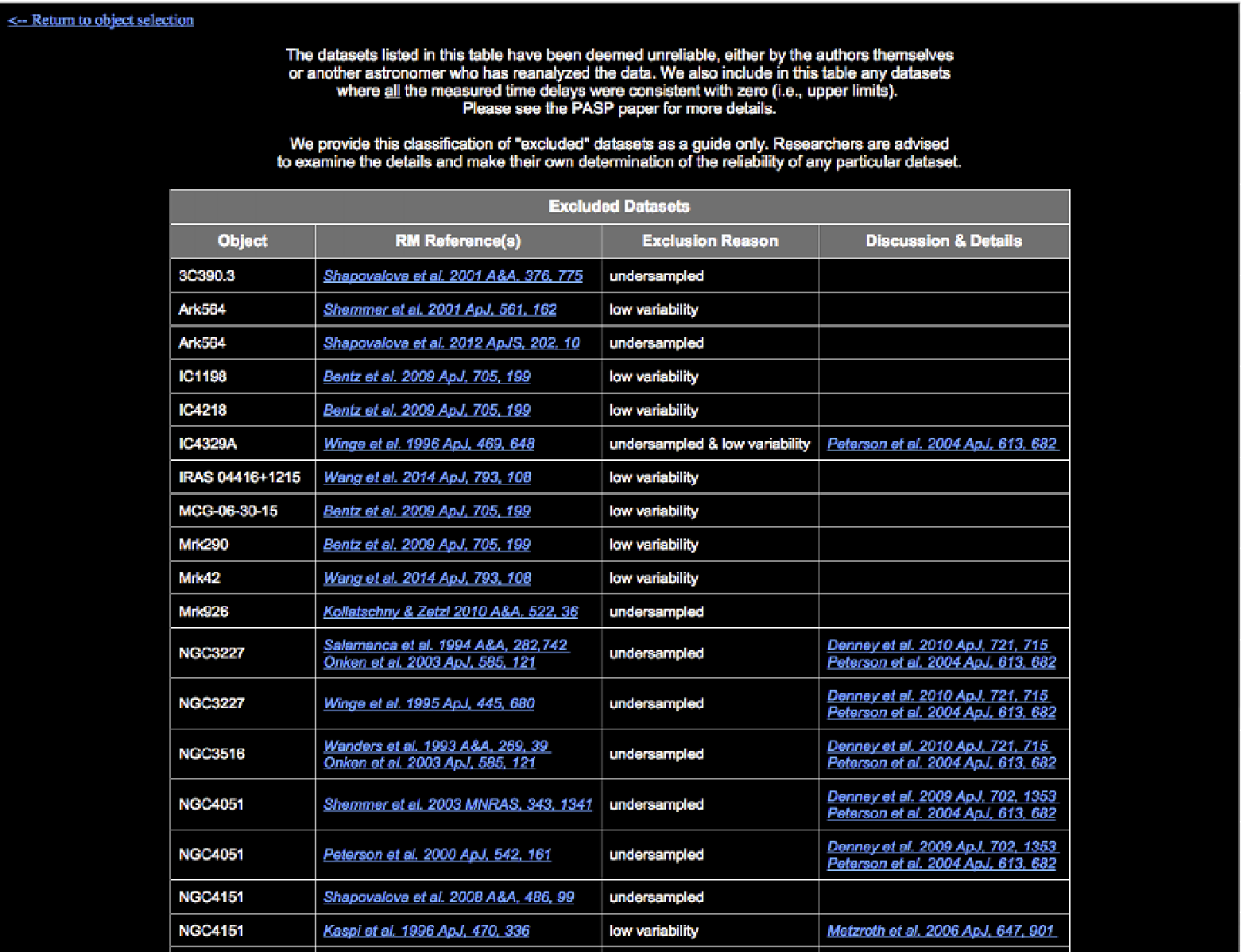}
\caption{The ``excluded data'' summary page.  The user may reach this
  page from the bottom of the main page, under the object table, or
  from a link on the individual page for any object with excluded
  data.  Twelve objects appear only in the excluded datasets table, as
  they have no reliable measurements for inclusion in the main
  database at this time.}
\label{fig:excluded}
\end{figure}

Excluded datasets are listed separately (Figure~\ref{fig:excluded})
and can be reached from a link on the front page below the main object
table, or from a link on a detailed object page if that object has
excluded data.  We do not tabulate the numbers for these datasets, as
they are untrustworthy and often unreported, but rather list the
object, the reference describing the monitoring campaign, the reason
the dataset was excluded, and any other reference that describes the
quality of the dataset in more detail (oftentimes, a reanalysis of the
original data, e.g., \citealt{peterson04}).

\section{Future Updates}

Reverberation mapping is a technique that has been exploited and
refined over the last $\sim 30$ years, and it continues to be a
standard tool for determination of black hole masses.  Thus, the AGN
Black Hole Mass Database will be a constantly-evolving compilation of
spectroscopic reverberation mapping results.  It is our intent to
continue updating the database as a service to the community.

To that end, we invite authors of papers with new reverberation
measurements to assist with this effort by sending copies of their
published or in press manuscripts to {\tt agnmass@astro.gsu.edu}.  New
results will be incorporated on a monthly basis once they have been
published by a peer-refereed journal and have their final
bibliographic identification.  The date of the last update to the
database can be found at the bottom of the main webpage.

We also plan to expand the database to include more information in the
future. In particular, we are currently compiling relevant details
about the AGN host galaxies, including the bulge stellar velocity
dispersion, the bulge magnitudes, and the inclination of the galaxy
disk (if a disk is present), and we plan to incorporate them into the
database in the near future.  Additional expansions are also likely to
follow as time and resources permit.

\section{Summary}

We have compiled a homogeneous collection of all published
reverberation mapping results and incorporated them into a SQL
database, hosted at Georgia State University.  We have also developed
a public web interface for the database, allowing users to quickly
identify objects with reverberation measurements and to identify the
most up-to-date determinations of \mbh\ from these measurements.
Additions to this compilation will be included as they enter the
published literature.  We also plan to incorporate other information
of interest, including detailed measurements of host-galaxy
characteristics, in the future.

\acknowledgements

We thank the anonymous referee for comments and suggestions that
helped to improve the presentation of the database and this
manuscript.  We thank Kelly Denney, Brad Peterson, Aaron Barth, and
Kate Grier for helpful feedback and discussions regarding the web
interface.  We thank Rachel Kuzio de Naray and Wei-Chun Jao for
helpful conversations and tips for working with SQL.  MCB gratefully
acknowledges support from the NSF through CAREER grant
AST-1253702. This research has made use of the NASA/IPAC Extragalactic
Database (NED) which is operated by the Jet Propulsion Laboratory,
California Institute of Technology, under contract with the National
Aeronautics and Space Administration.

\bibliographystyle{apj} 

\clearpage

\end{document}